\documentclass[12pt]{iopart}

\usepackage{graphicx} 
\usepackage{epstopdf}
\usepackage{amssymb}
\usepackage{amsbsy}
\usepackage{mathenv}
\usepackage{color}
\usepackage{ulem}

\newcommand{\YbTO}{{\rm Yb_{2}Ti_{2}O_{7}}}
\newcommand{\ATO}[1]{{\rm #1_{2}Ti_{2}O_{7}}}

\newcommand{\CW}{\theta_{\rm CW}}

\newcommand{\crysvec}[1]{[#1]}
\newcommand{\chiL}{\chi_{a}}
\newcommand{\titanate}[1]{\rm #1_{2}Ti_{2}O_{7}}

\newcommand{\chipar}{\chi_{\parallel}}
\newcommand{\chiperp}{\chi_{\perp}}
\newcommand{\tfrac}[2]{\textstyle\frac{#1}{#2}}

\begin{document}
\title[Local Susceptibility of the Yb$_2$Ti$_2$O$_7$ Rare Earth Pyrochlore]
{Local Susceptibility of the Yb$_2$Ti$_2$O$_7$ Rare Earth Pyrochlore Computed from 
a Hamiltonian with Anisotropic Exchange} 
\author{J D Thompson$^1$, P A McClarty$^1$, and M J P Gingras$^{1,2}$}
\address{${}^{1}$Department of Physics and Astronomy, University of Waterloo, Waterloo, ON, N2L 3G1, Canada.}
\address{${}^{2}$Canadian Institute for Advanced Research, 180 Dundas Street West, Toronto, Ontario M5G 1Z8, Canada}

\begin{abstract}
The rare earth pyrochlore magnet $\YbTO$ is among a handful of materials that
  apparently exhibit no long range order 
down to the lowest explored temperatures and well below the Curie-Weiss temperature. 
Paramagnetic neutron scattering 
on a single crystal sample has revealed the presence of anisotropic correlations and 
recent work has led to the proposal 
of a detailed microscopic Hamiltonian for this material involving significantly 
anisotropic exchange. In this article, 
we compute the local sublattice susceptibility of $\YbTO$ from the proposed model 
and compare with the measurements of
 Cao and coworkers \cite{Cao}, finding quite good agreement. 
In contrast, a model with only isotropic exchange and 
long range magnetostatic dipoles gives rise to a local susceptiblity 
that is inconsistent with the data.
\end{abstract}

\section{Introduction}

The pyrochlore lattice of corner-sharing tetrahedra is a classic example of a
lattice prone to geometric frustration, the inability of interacting magnetic
moments on such a lattice to simultaneously minimize all of the pairwise
interactions in the system. In the rare earth oxides, {\it A}$_2${\it B}$_2$O$_7$,
where {\it A} is a trivalent rare earth ion (Ho, Dy, Tb, Gd, Yb) or yttrium
(Y) and {\it B} is a tetravalent transition metal ion (Ti, Sn, Mo, Mn), both
{\it A} and {\it B} reside on distinct interpenetrating pyrochlore lattices. 
Such systems have been the subject of intense
research for the past 10 years as they are known to manifest a wide variety of
exciting collective phenomena \cite{GGG} including spin ices ($\ATO{Dy}$, $\ATO{Ho}$)
\cite{Bramwell-Science}, spin liquids ($\ATO{Tb}$) \cite{Gardner-TTO-PRL},
spin glasses (Y$_2$Mo$_2$O$_7$) \cite{Gingras-YMoO-PRL}, and LRO with
persistent spin dynamics (Gd$_2$Sn$_2$O$_7$, Er$_2$Ti$_2$O$_7$) \cite{Gd2Sn2O7,ETO}.

To better understand the many complex phenomena observed in magnetic
pyrochlore oxides, knowledge of the microscopic interactions present in these materials is necessary.
The pyrochlore lattice can be described as an FCC Bravais lattice with a
tetrahedral basis at each FCC site. The magnetic rare earth ($A^{3+}$) ions reside on the pyrochlore
lattice and are subject to a crystal field (CF) which introduces a magnetic
anisotropy with respect to the local $\langle 111\rangle $ axes that is reflected
in the magnetic susceptibility on each of the four sublattices. Recently,
determination of the local susceptibility, $\chiL$, via polarized neutron
scattering has been brought to bear on the rare earth pyrochlores including
$\titanate{Tb}$, $\titanate{Ho}$, $\titanate{Er}$, and $\titanate{Yb}$
\cite{Cao}. These measurements have been used to analyse the single ion
anisotropy and the nature of the magnetic interactions \cite{Cao,Malkin}. In
particular, despite concerns about some details of the method with which the
theoretical analysis of $\chiL$ is carried out in Refs.~\cite{Cao,Malkin}, which
we discuss later on in the Appendix, both of these works arrive at the tentative conclusion
that in most of these materials, isotropic exchange interactions are
inadequate to account for the $\chiL$ measurements and that, instead, the
interactions appear to some extent anisotropic. The presence of anisotropic
exchange in the pyrochlore oxides provides a further clue to the vehicle for
the richness of phenomena observed in rare earth frustrated magnets \cite{GGG}.  
Indeed, recent theoretical work has exposed large
anisotropic superexchange in pyrochlore oxide
 materials \cite{Onoda-1,Onoda-2}.

In this paper, we focus on local susceptibility measurements on one material,
$\YbTO$ \cite{Cao}, which is remarkable for exhibiting a phase transition at
$T_c \approx 240$ mK, albeit with no signs of long range magnetic order at lower
temperatures \cite{Hodges-YbTO-PRL}. The nature of the low temperature phase
in $\YbTO$ is currently somewhat controversial. Reference ~\cite{Yasui} reports the
onset of hysteresis in the vicinity of $240$ mK and the presence of magnetic
Bragg peaks at lower temperatures. Other studies, however, have found no
evidence of magnetic reflections \cite{Hodges-YbTO-PRL,Gardner-YbTO,Ross} in
zero applied field. 
Reference ~\cite{Hodges-YbTO-PRL} presents muon spin relaxation data revealing that the spin
fluctutation rate drops sharply at about 240 mK, but remains nonzero and temperature independent
well below $T_c$,  perhaps indicating a spin gas to spin liquid type phase transition.
Peculiarly, the collective paramagnetic state
exhibits strongly anisotropic rod-like correlations in the diffuse neutron
scattering which have been conjectured to arise from a weak decoupling of the
kagome layers in the pyrochlore lattice \cite{Ross}. We have
investigated the origin of the correlations seen in neutron scattering by 
using a random phase approximation calculation 
of the paramagnetic neutron scattering intensity pattern and found
a set of interactions that produce a theoretical
scattering pattern that is in very good agreement with the
experimental data \cite{YbTO}. The foremost characteristic of these
interactions is that they are strongly anisotropic. The purposes of 
the present work
are (i) to provide a comparison of the predictions of our 
microscopic Hamiltonian \cite{YbTO}
with the local susceptibility ($\chi_a$)  data of Ref.~\cite{Cao} and (ii) to provide a discussion
of the sensitivity of $\chi_a$ to the nature of the interactions and compare
with the results of Refs.~\cite{Cao} and~\cite{Malkin}. We show that our model
can successfully account for the observed temperature dependence of $\chi_a$
without further adjustment to the exchange couplings determined in
Ref.~\cite{YbTO}.

\section{Anisotropic Exchange Model}
\label{sec:model}
In this section, we 
provide a description of a candidate theory for the magnetism of $\YbTO$ introduced in Ref.~\cite{YbTO}. 
The magnetic Yb$^{3+}$ ion has electronic configuration $^2$F$_{7/2}$, such
that ${\rm J}=7/2$ and Land\'{e} factor $g_{\rm J}=8/7$. The nearest neighbour
distance between Yb$^{3+}$ ions is $r_{\rm nn}=(a/4){\sqrt {2}}$, where
$a=10.026$ \AA{} is the size of the conventional cubic unit cell
\cite{Gardner-YbTO}. The single ion crystal field interaction $H_{\rm cf}$ 
is the largest magnetic energy scale.
 The form of $H_{\rm cf}$ is fixed by the symmetry of the Yb$^{3+}$ environment. 
To specify $H_{\rm cf}$ completely, 
we require the values of the six crystal field (CF) parameters. We use two independent
 sets of CF parameters determined in Refs.~\cite{Cao,Hodges-YbTO-JPC}. 
The magnetic Hamiltonian is  ${H} = H_{\rm cf} + H_{\rm int}$, 
which includes both the CF and spin-spin interactions
of the form $H_{\rm int}=H_{\rm ex}+H_{\rm dip}$ which is composed of exchange, $H_{\rm ex}$, 
and long range magnetostatic dipolar interactions  
$H_{\rm dip} = \sum_{i>j;a,b} \frac{D (r_{\rm nn})^3}
{\mid  \mathbf{R}_{ij}^{ab}\mid^{3}} 
[ \mathbf{{J}}_{i}^{a} \cdot \mathbf{{J}}_{j}^{b} 
-3 (\mathbf{{J}}_{i}^{a} \cdot \hat{\mathbf{R}}_{ij}^{ab} )
 (\mathbf{{J}}_{j}^{b} \cdot \hat{\mathbf{R}}_{ij}^{ab} ) ]$
 with coupling
$D=\frac
{
 {\mu_{0}(g_{\rm J} \mu_{\rm B})^2}}
{  {4\pi {r_{\rm nn}}^3}
}
 \approx 0.01848$ K.
$H_{\rm ex}$ contains all nearest neighbour exchange
interactions, $\mathcal{J}_e$, that respect lattice symmetries. 
There are four such nearest neighbour interactions \cite{1742-6596-145-1-012032}:
$H_{\rm Ising}=-\mathcal{J}_{\rm Ising}
\sum_{<i,j>;a,b}\left(\mathbf{{J}}_{i}^{a}
\cdot\mathbf{\hat{z}}^{a}\right)\left(\mathbf{{J}}_{j}^{b}\cdot\mathbf{
\hat{z}}^{b}\right)$, 
which couples the local $\left[111\right]$ ${\hat z}$ components of 
$\mathbf{J}$ (shown in Fig.~\ref{fig:1}(a)), 
$H_{\rm iso}=
-\mathcal{J}_{\rm iso}\sum_{<i,j>;a,b}
\mathbf{{J}}_{i}^{a} \cdot \mathbf{{J}}_{j}^{b}$,  the standard isotropic exchange, 
$H_{\rm pd} =-\mathcal{J}_{\rm pd}\sum_{<i,j>;a,b} 
[ \mathbf{{J}}_{i}^{a} \cdot \mathbf{{J}}_{j}^{b} - 3 (\mathbf{{J}}_{i}^{a} \cdot
\hat{\mathbf{R}}_{ij}^{ab} ) (\mathbf{{J}}_{j}^{b}
 \cdot \hat{\mathbf{R}}_{ij}^{ab} ) ]$, 
a pseudo-dipolar  interaction of exchange origin and not
part of $H_{\rm dip}$ and, finally, 
$H_{\rm DM}=-\mathcal{J}_{\rm DM}\sum_{<i,j>;a,b}\boldsymbol{\Omega}^{a,b}_{{\rm DM}} 
\cdot \left(\mathbf{{J}}_{i}^{a} \times \mathbf{{J}}_{j}^{b}\right)$, 
the Dzyaloshinskii-Moriya (DM) interaction~\cite{Canals2}. 
 In all of these terms, 
$\mathbf{J}_{i}^{a}$
 denotes the angular momentum of the Yb$^{3+}$
 located at lattice $\mathbf{R}_{i}^{a}$
 (FCC lattice site $i$ 
 and tetrahedral sublattice site $a$) and
 $\hat{\mathbf{R}}_{ij}^{ab}$ is a unit vector directed 
along $\mathbf{R}_{j}^{b}-\mathbf{R}_{i}^{a}$.

In our recent work \cite{YbTO}, we were able to determine the values of
$\mathcal{J}_e\equiv\{\mathcal{J}_{\rm iso},\mathcal{J}_{\rm
  Ising},\mathcal{J}_{\rm pd},\mathcal{J}_{\rm DM} \}$. We accomplished this
by fitting the diffuse paramagnetic neutron scattering measurements to neutron scattering
patterns generated using the random phase approximation \cite{YbTO}. The
fitting was done by treating the four exchange couplings, $\mathcal{J}_e$, as
free parameters, and determining their optimal values using simulated
annealing to minimize the difference between the model and experimental
scattering patterns, along with a term to fix $\CW$ to the value reported in Ref.~\cite{Hodges-YbTO-JPC}
 of $\CW = 0.75$ K \cite{Note}. 
The values of $\mathcal{J}_e$ we extracted are \cite{YbTO}:
\begin{equation}
 \mathcal{J}_{\rm Ising}  =  0.81~{\rm K} \hspace{5pt}
 \mathcal{J}_{\rm iso}  = 0.22~{\rm K} \hspace{5pt}
 \mathcal{J}_{\rm pd}  =  -0.29~{\rm K} \hspace{5pt}
 \mathcal{J}_{\rm DM}  =  -0.27~{\rm K} \label{eqn:1}
\end{equation} 
when using the CF parameters of \cite{Hodges-YbTO-JPC}, and
\begin{equation}
 \mathcal{J}_{\rm Ising}  = 0.76~{\rm K} \hspace{5pt}
 \mathcal{J}_{\rm iso}  = 0.18~{\rm K} \hspace{5pt}
 \mathcal{J}_{\rm pd}  = -0.26~{\rm K} \hspace{5pt}
 \mathcal{J}_{\rm DM} = -0.25~{\rm K} \label{eqn:2}
\end{equation} 
 when using the CF parameters of \cite{Cao}. 
The values of $\mathcal{J}_{\rm DM}$ 
in both of these sets of 
interactions are large, even compared to values found in other insulating pyrochlore materials 
based on magnetic d-shell transition metal ions that display significant
DM interactions \cite{Onose,Chern}. However,
 it has recently been shown that large DM interactions 
are allowed in Yb$_2$Ti$_2$O$_7$ due to 
antisymmetric contributions to the superexchange 
between Yb$^{3+}$ ions mediated via the oxygen ions at the O1 site
(i.e. in the middle of each tetrahedron) \cite{Onoda-2}.

\section{Local Susceptibility}
\label{sec:theory}

We briefly outline the method of Gukasov and Brown \cite{Gukasov} for determining 
the components of the 
site susceptibility tensor  from
 polarized neutron scattering. In the presence of a magnetic field, the induced moment on sublattice 
$a$ at fcc site $\mathbf{R}_{i}$ is given by $\mathbf{M}_{a}(\mathbf{R}_{i})$ where we assume that 
there is no contamination from $\mathbf{k}\neq 0$ magnetic structures. 
The magnetic Bragg intensity $I({\mathbf q})$ depends on 
$\mathbf{{M}}(\mathbf{q})$, 
the magnetic moment in reciprocal space, and is given by
\begin{eqnarray*}
I(\mathbf{q}) \propto \left( \sum_{\mathbf{G}}
\delta(\mathbf{q}-\mathbf{G})  \right)\left( \bar{b}^{2} 
+ \bar{b}\mathbf{P}_{0}\cdot \mathbf{T}(\mathbf{q}) 
+ \frac{1}{4} \mathbf{T}(\mathbf{q})\cdot \mathbf{T}(\mathbf{q})  \right)  
\end{eqnarray*}
where $\mathbf{T}(\mathbf{q})=\frac{1}{2}aF(\mathbf{q})[\mathbf{M}(\mathbf{q})
  - \mathbf{\hat{q}}(\mathbf{M}(\mathbf{q})\cdot\mathbf{\hat{q}})]$ and both
$\bar{b}$ and $a$ are constants. Since the neutron intensity depends on the
neutron polarization vector $\mathbf{P}_{0}$, measurements of the
spin-non-flip and spin-flip intensity for different Bragg peaks provide
information about the moments on each sublattice.  

The site susceptibility $\chi_{a}$ is defined as
$M_{a}^{\alpha}(\mathbf{R}_{i})=\bar{\chi}_{a}^{\alpha\beta}h^{\beta}$ where
$h^{\beta}$ is the applied magnetic field and $\alpha,\beta$ are components in
the crystallographic frame, with the bar over $\chi_{a}$ indicating the choice
of the crystallographic frame. In the refinement of the moments from the
neutron data,
$\bar{\chi}_{a}$  on sublattice $1$ (at the fcc sites) was taken to
have two components with all diagonal elements equal and all off-diagonal
elements equal in the crystallographic frame. In the experiment \cite{Cao} on
$\YbTO$, the applied magnetic field was $1$ T in the $[110]$ direction, so the
experiment was sensitive to components $\bar{\chi}_{1}^{\alpha\beta}$ with
$\alpha=1,2$. Figure~\ref{fig:1}(a) shows the tetrahedral basis with the
choice of sublattice labelling as well as the local Ising $\langle 111
\rangle$ directions and the $[110]$ orientation of the applied magnetic
field. 

We computed the moments and hence the site susceptibility within a local mean field theory 
by computing the single ion spectra self-consistently in the presence of a magnetic field. 
Each bilinear spin operator ${\rm J}_{i}^{\alpha}{\rm J}_{j}^{\beta}$, with spin components 
$\alpha $ and $\beta$, in the interaction Hamiltonian $H_{\rm int}$ was written as
\begin{eqnarray*}
  {\rm J}^{\alpha}_{i}{\rm J}^{\beta}_{j} \rightarrow {\rm J}_{i}^{\alpha}\langle {\rm J}_{j}^{\beta} \rangle
 +  \langle {\rm J}_{i}^{\alpha}\rangle {\rm J}_{j}^{\beta}  
- \langle {\rm J}_{i}^{\alpha}\rangle \langle {\rm J}_{j}^{\beta} \rangle
 + \left({\rm J}_{i}^{\alpha} -  \langle {\rm J}_{i}^{\alpha}\rangle \right)({\rm J}_{j}^{\beta} 
-  \langle {\rm J}_{j}^{\beta}\rangle )  
\end{eqnarray*}
and the last term on the right-hand side - the fluctuation term - 
was dropped to obtain a decoupled mean field Hamiltonian 
$H_{\rm int,MF}\equiv \sum_{\rm i} H_{\rm int,MF}({\rm i})$ for each  site $\rm i$. The single ion wavefunctions 
$\vert \nu\rangle$, at site $\rm i$, were obtained from $H_{\rm MF}\equiv H_{\rm
  cf} + H_{\rm Z}+H_{\rm int,MF}$~\cite{Cao,Hodges-YbTO-PRL} 
where $H_{\rm Z}$   is the Zeeman interaction term, 
$H_{\rm Z}=-\mu_{\rm B}g_{\rm J}\sum_i {\mathbf  J}^{\alpha}\cdot{\mathbf B}$
 by solving for the moments $\langle {\rm  J}_{i}^{\alpha}\rangle$. 
 The infinite lattice sum of the dipolar interaction
  was computed using Ewald summation~\cite{Gingras-MFT}. From a set of
  randomly chosen initial moments, $H_{\rm MF}$ was diagonalized on each site
  of a cubic unit cell with periodic boundary conditions and the resulting
  spectrum, $\vert \nu, {\rm i}\rangle$, for sites $i=1,\ldots,16$ and
  $\nu=1,\ldots,2{\rm J}+1$ used to obtain a new set of moments $\langle {\rm
  J}_{i}^{\alpha}\rangle = Z^{-1}{\rm Tr}\{{\rm J}_{i}^{\alpha} \exp(-\beta
  H_{\rm MF}({\rm i})) \}$ where $Z={\rm Tr}\{ \exp(-\beta H_{\rm MF}({\rm
  i}))\}$. The process of successive diagonalization and computation of the
  moments was iterated until convergence was reached. The process was repeated
  for each temperature. The use of a single cubic unit cell
 for the diagonalization is more than sufficient since it was found in Ref. \cite{YbTO} 
that the interaction Hamiltonian $H_{\rm int}$ with parameters from either
 Eq. \ref{eqn:1} or \ref{eqn:2} 
leads to $q=0$ ordering, signifying an identical moment configuration on each
tetrahedral primitive unit cell.

  From the moments obtained within MF theory, we computed the single ion
  susceptibility $\chi_{a}$ for each sublattice $a=1,2,3,4$ - each independent
  (primitive) tetrahedron being identical. In common with
  Refs.~\cite{Cao,Malkin}, we present components of the susceptibility
  parallel ($\chi_{\parallel}$) and perpendicular ($\chi_{\perp}$) to the
  local Ising directions. These components were computed by introducing
  rotation matrices $u_{a}^{\alpha\beta}$ from the global to the local frame
  with $\hat{z}_{a}$ axes along the $\langle 111\rangle$
  directions. Transforming $\bar{\chi}_{a}$ to the local frame
  $\chi_{a}^{\alpha\beta}=
  u_{a}^{\alpha\gamma}u_{a}^{\beta\delta}\bar{\chi}_{a}^{\gamma\delta}$, we
  extract $\chi_{a,\parallel}\equiv \chi_{a}^{zz}$. Our choice of local frame
  is such that the $[110]$ field has vanishing local $y$ component for
  sublattices $1$ and $2$ (along the $\alpha$ chains) and vanishing local $x$
  and $z$ components for sublattices $3$ and $4$ (along the $\beta$
  chains). Hence, for sublattices $1$ and $2$, with an applied $[110]$ field,
  we take $\chi_{a,\perp} = \chi^{xx}$ and for sublattices $3$ and $4$,
  $\chi_{a,\perp}=\chi^{yy}$.  
  
\section{Results}

\begin{figure}[htb!]
\begin{center}
\includegraphics[scale = 0.7]{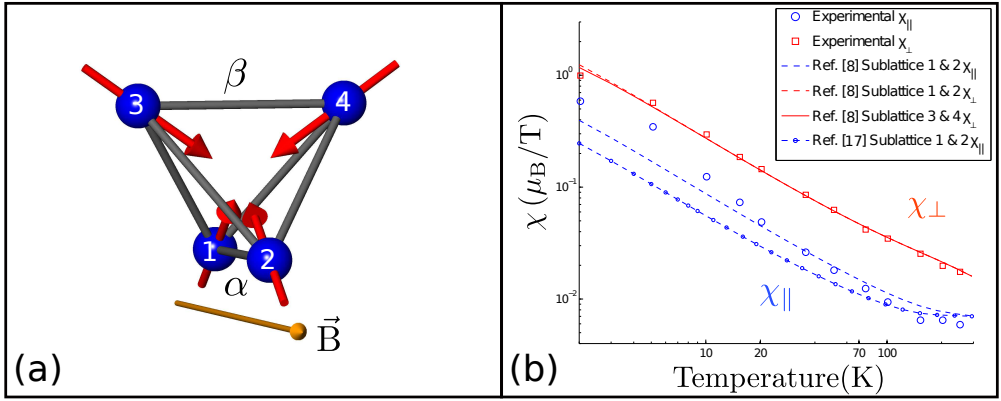}
\caption{(a) Tetrahedral basis of the pyrochlore structure showing the four anisotropy axes 
as red arrows and the applied magnetic field $\mathbf{B}$ orientation (orange arrow). 
Sublattices $1$ and $2$ make up the $\alpha$ chains and sublattices $3$ and $4$, the $\beta$ chains. 
(b) Local susceptibility (open symbols) as reported in Ref.~\cite{Cao} in a $1$ T field along $\crysvec{110}$, 
along with single ion local susceptibilities computed using the crystal fields of Ref.~\cite{Cao} 
(dashed and solid lines) and Ref.~\cite{Hodges-YbTO-JPC} (dot-dash line - the perpendicular part 
having been omitted as it almost coincides with the Ref.~\cite{Cao} crystal field result) in 
the same field, but without including the $\mathcal{J}_e$ exchange interactions.}
\label{fig:1}
\end{center}
\end{figure}

Fig.~\ref{fig:1}(b) shows the single ion $\chiL=\{\chi_{\parallel},\chi_{\perp}\}$, computed only from the Zeeman
 term and CF Hamiltonian using the CF parameters of Refs.~\cite{Cao,Hodges-YbTO-JPC}, and the experimentally 
determined values of $\chiL$ \cite{Cao}. Both sets of CF parameters give rise to a $\chiperp$ that fits the
 experimental data well over the whole temperature range even in the absence of interactions except for 
the lowest temperature point, $T=2$ K, where the experimental $\chiperp$ falls below the crystal field prediction. 
However, the $\chipar$ calculations for the two sets of CF parameters are significantly different from one another 
and from the experiment. Whereas, the parameters of Ref.~\cite{Cao} give a $\chipar$ that fits reasonably well 
between $20$ K and $70$ K, deviating from experiment at the highest and lowest measured temperatures, 
those of Ref.~\cite{Hodges-YbTO-JPC} do not fit well over most of the temperature range.  
The addition of anisotropic exchange of the form given in Eq.~(\ref{eqn:1}), does not improve 
the fit to experiment of $\chipar$ computed using the CF parameters of Ref.~\cite{Hodges-YbTO-JPC}.  
Because of the poor fit to $\chipar$, even when anisotropic exchange is included, we do not use the 
CF parameters of Ref.~\cite{Hodges-YbTO-JPC} any further in this work despite their fair success in
 reproducing paramagnetic neutron scattering measurements in Ref.~\cite{YbTO}. 
Another point to be drawn from Fig.~\ref{fig:1}(b) is that there is a difference between 
the local susceptibilities for the $\alpha$ and $\beta$ chains. In particular, $\chipar$ 
is not probed for the $3$ and $4$ sublattices because the $[110]$ field does not induce a 
moment in the local $[111]$ Ising direction on these sublattices. Also, $\chiperp$, which is 
probed on all four sublattices, is different for the $\alpha$ and $\beta$ chains.
 As the temperature is raised, the cubic symmetry between different sublattices is partially restored.

\begin{figure}[htbp]
\begin{center}
\includegraphics[scale=0.8]{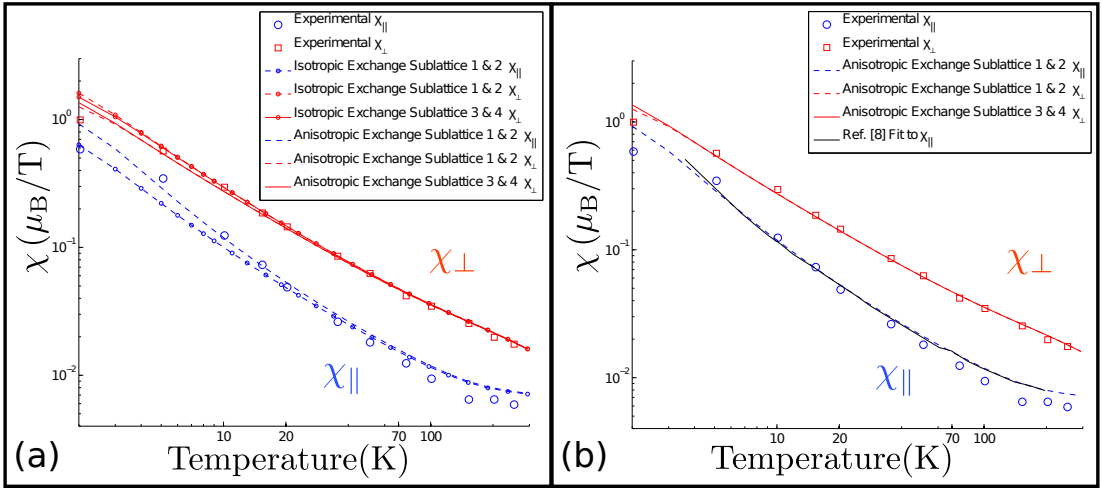}
\caption{(a): $\chiL$ as reported by \cite{Cao} (open squares and circles). Also shown is $\chiL$, 
for all four magnetically inequivalent sublattices, computed within local mean field theory 
with our anisotropic exchange model with long range dipoles obtained from neutron scattering 
data in Ref.~\cite{YbTO}. The temperature range is $2$ K to $300$ K. Also shown is the
prediction of a model with isotropic exchange interactions and long ranged dipoles. 
(b): A comparison of our anisotropic exchange model with the prediction of the model of Ref.~\cite{Cao}.}
\label{fig:2}
\end{center}
\end{figure}

When interactions are included in the computation of $\chiL$ as described in
  Section~\ref{sec:theory}, we obtain the results shown in
  Fig.~\ref{fig:2}. Figure \ref{fig:2}(a) shows the local susceptibility,
  computed for two different models over the temperature range $2$ K to $300$
  K, together with the experimental data of Ref.~\cite{Cao}. The inclusion of
  interactions does not significantly affect $\chiperp$ so, for these results,
  we merely observe that the splitting between $\alpha$ and $\beta$ chains is
  present once again because of the symmetry breaking field. The first model
  to be compared with the experimental $\chipar$ data is one with long ranged
  dipoles and purely isotropic exchange ($\mathcal{J}_{\rm
  Ising}=\mathcal{J}_{\rm
  pd}=\mathcal{J}_{\rm DM} = 0$), with a coupling $\mathcal{J}_{\rm
  iso}=-0.06$ K constrained by the Curie-Weiss temperature which is taken to
  be $\CW=0.75$ K, within the experimental error margins given above
  \cite{YbTO} (blue dot-dash line). The second model is the candidate
  anisotropic exchange model described in Section~\ref{sec:model} and
  Ref.~\cite{YbTO} (blue dashed line). The $\chi_{\perp}$ results (red lines)
  do not strongly distinguish between the two models. At low temperatures,
  $T\lesssim 30$ K, where the effect of interactions produces significant
  deviations of $\chipar$ from the noninteracting case, the isotropic exchange
  model does not fit the experimental data as well as the anisotropic exchange
  model, except at the lowest temperature point ($2$ K). However, neither model
  captures the $2$ K point for $\chiperp$.

Fig.~\ref{fig:2}(b) shows once again the experimental data and the anisotropic
model predictions for $\chiL$ as well as $\chiL$ computed using the molecular
field method of Ref.~\cite{Cao}. The fit in  Ref.~\cite{Cao} of  $\chipar$ is
only provided down to approximately $3.5$ K in that work (and no results for
$\chiperp$ for the single ion mean field case are reported in that reference).
For the temperature range over which the models can be compared, our
anisotropic exchange model 
describes the experimental data as well as the model of Ref.~\cite{Cao}.
However, our model achieves this while correctly
preserving lattice symmetries, unlike the model of Ref. [8],
and therefore corresponds to a physically acceptable microscopic model.
We discuss further the comparison of our results with those
of Refs.~\cite{Cao,Malkin} in the Appendix.

At temperatures $T\gg \theta_{\rm CW}$, interactions are expected to have
little effect on $\chiL$ which is borne out by our calculations. Therefore the
differences between the $\chiL$ calculations and experimental results at high
temperatures ($T >100$ K) can be attributed to an incorrect spectral
decomposition of the excited crystal field states \cite{Cao,Hodges-YbTO-JPC}.  

\section{Discussion}

We have investigated the local susceptibility computed from the anisotropic
exchange model described in Section~\ref{sec:model} for $\YbTO$. The model was
obtained from an {\it independent} previous fit to the paramagnetic neutron
scattering \cite{YbTO}. The calculations presented here fit the $\chiL$
experimental data well with no adjustable parameters over much of the
experimental temperature range ($2$ K to $250$) K. There are, however,
discrepancies in the fit above approximately $100$ K, presumably due to the fact that
the available CF parameters do not capture the correct 
wavefunctions for the excited CF energy levels.
 The fit is also imperfect at the lowest observed temperature of
$2$ K where the experimental $\chiL$ susceptibility flattens out while it
does not in the mean field treatment of the Hamiltonian 
$H=H_{\rm cf} + H_{\rm int} + H_{\rm Z}$.

At high temperatures $T\gg \theta_{\rm CW}$,
$\chiL$ is insensitive to the interactions and the quality of fit is
determined by the CF Hamiltonian and the Zeeman term. Fig.~\ref{fig:1}
illustrates that the crystal field provides an adequate fit to $\chipar$ from $100$ K down
to about $30$ K and
to $\chiperp$ over the entire range of temperatures. Figure~\ref{fig:2} shows
that the quality of fit for $T>30$ K does not change with the
inclusion of interactions. Therefore, the only nontrivial test of our model
comes from the five lowest temperature points ($T=2$ K, $5$ K, $10$ K, $15$ K,
$20$ K) in $\chipar$ where the noninteracting Hamiltonian,
$H=H_{\rm cf} +  H_{\rm Z}$, 
 gives a $\chipar$ that falls significantly below the experimental result
(see Fig.~\ref{fig:1}(b)). The anisotropic exchange model determined in 
Ref.~\cite{YbTO} provides a good description of $\chipar$, 
whereas isotropic exchange interactions do not -- this is the main
conclusion of this work. Although the relevant test of our model comes from
only a few (five) data points, it achieves the goals of providing a {\it consistency check} of our model
determined previously on the basis of diffuse paramagnetic neutron scattering
\cite{YbTO} while also establishing that anisotropy in the exchange is indeed
an intrinsic ingredient of the spin Hamiltonian of $\YbTO$.


Refs.~\cite{Cao} and \cite{Malkin} both extract exchange parameters from the 
local susceptibility data, and present fits to the local susceptibility that
 result from these exchange parameters.  As stated above,
we find that this method for 
extracting exchange parameters is under constrained.
In the Appendix, we describe in some detail the approaches taken in 
these two papers and compare with our own results.

\section{Conclusion}

In this work we have shown that the anisotropic exchange model extracted from
neutron scattering measurements in a previous work \cite{YbTO} produces a good
fit to both local and bulk susceptibility measurements on the rare earth
pyrochlore material $\YbTO$ for temperatures in the range $T\lesssim 100$ K.
This model has four nonadjustable and 
a priori prescribed bilinear exchange couplings which were
extracted from neutron scattering data \cite{YbTO}. The fit of our model to
the experimental measurements of $\chiL$ and the bulk susceptibility provides
further {\it independent} evidence for the correctness of our anisotropic exchange model \cite{YbTO}.

We also found that the origin of the discrepancy between the experiment
and our model at high temperatures, and indirectly, a similar failure to fit
the local susceptibility
$\chiL$ also at $T \gtrsim 10^2$ K in Ref.~\cite{Cao}, lies in the excited
CF states obtained from the CF parameters of Refs.~\cite{Cao,Hodges-YbTO-JPC}. 
Further experimental work would be required to probe more accurately the spectral
decomposition of the excited states of the crystal field in order to better determine their structure.
That said, we do not expect that the corrections to the spectral decomposition
of those excited states would, as a byproduct, necessitate revisiting the
value of the anisotropic exchange parameters ${\mathcal J}_e$.
We are thus confident, combining the
results presented in this paper and those of Ref.~\cite{YbTO}, that we have in
hand the correct microscopic Hamiltonian of $\YbTO$ that includes all
symmetry-allowed anisotropic bilinear nearest-neighbour couplings.  One may
therefore aim to proceed to unravel the nature of the low temperature
($T\lesssim2$ K) state of this exotic material.
In this context, it is worth commenting on the validity of 
the mean field theory treatement of our model at
low temperatures ($T\lesssim 2$ K) and/or in weak field.
Mean field theory predicts a transition to a $q=0$ 
structure at approximately 1.1 K \cite{YbTO}. This contrasts
with the experimental $T_c=240$ mK \cite{Hodges-YbTO-PRL}.
As usual, mean field calculations neglect the effects of
both thermal and quantum fluctuations.
These could be fairly large in this system and greatly  
decrease the critical temperature. The role of
thermal and quantum fluctuations in the proposed Hamiltonian
$H=H_{\rm int}+H_{\rm cf}$ for Yb$_2$Ti$_2$O$_7$
remains to be investigated.
To the best of our knowledge, and
notwithstanding the difficulties with mean field theory,
the sharp specific heat feature signaling
a phase transition in Yb$_2$Ti$_2$O$_7$
has only been reported in powder samples.
It would be useful to confirm the existence of this
feature in 
single crystals.


Two interesting observations that
arise from this work are that 
(i) the  isotropic exchange ${\mathcal J}_{\rm iso}$ is
ferromagnetic (positive)  in Yb$_2$Ti$_2$O$_7$ 
while it is found to be antiferromagnetic (negative) for the other
${\it A}_2$Ti$_2$O$_7$ compounds and (ii) there appears to exist
large antisymmetric anisotropic exchange, described above in terms 
of the form of a Dzyaloshinskii-Moriya (DM) interaction. Very recent theoretical
work \cite{Onoda-2} has found that large anisotropic 
exchange, including large DM interactions, can in principle
arise in Yb$_2$Ti$_2$O$_7$. It would be interesting to investigate this
issue further.

The conclusion of this work, that the magnetic exchange interactions in $\YbTO$
are anisotropic, hints at the possibility of significant multipolar
interactions \cite{Onoda-1,Santini} in this material and, perhaps, in other rare earth
pyrochlores.  That said, since the crystal field gap is much larger in $\YbTO$ than the
scale of the interactions, 
anisotropic bilinear couplings between the ${\mathbf J}_i$ operators as
as well as multipolar couplings between the Yb$^{3+}$ moments
manifest themselves, at low energies, in the form of generally anisotropic
bilinear exchange couplings between effective pseudospins one-half \cite{Onoda-2}.

\ack

We thank  Shigeki Onoda for useful and stimulating discussions.
This research was funded by the NSERC 
of Canada and the Canada Research Chair program (M. G., Tier I).

\appendix
\section{Previous works on local susceptibility in $\YbTO$}
\label{app:a}

In this Appendix we discuss the anisotropic exchange fits of Refs.~\cite{Cao}, for $\YbTO$, and
\cite{Malkin} for various rare earth pyrochlore titanates.

The fit to $\chiL$ presented in Ref.~\cite{Cao} and reproduced in Fig.~\ref{fig:2}b is based on a 
self-consistent mean field theory that considers  a single sublattice. 
The moment $\langle {\mathbf J}\rangle$, 
at a temperature $T$, was computed from a mean field Hamiltonian consisting of the crystal field, 
a Zeeman term, and a mean field term arising from the interactions of the form 
$-g_J\mu_B\lambda^{C}_\alpha\langle J_\alpha\rangle$, where $\alpha$ specifies the components
 $\parallel$ and $\perp$, the directions parallel and perpendicular to the local $[ 111 ]$ direction. 
Solving for the moment, the two couplings $\lambda_{\parallel}$ and $\lambda_{\perp}$ 
were determined from a fit to $\chiL$. 
Since this model fails to take into account the sublattice structure of the pyrochlore lattice, 
the ensuing fit is unlikely to correctly reflect  
the underlying microscopic couplings in $\YbTO$. 

The model of Ref.~\cite{Malkin} includes the crystal field anisotropy as well as nearest neighbour 
exchange couplings and the long range dipole interaction (treated via an Ewald summation). 
The model correctly handles the sublattice structure of the pyrochlore lattice. 
The model includes, at the outset, three exchange couplings which respect the lattice 
symmetries and which are denoted: $\lambda^{M}_{\perp,1}$, $\lambda^{M}_{\perp,2}$ and $\lambda^{M}_\parallel$. 
These can be expressed in terms of the couplings $\mathcal{J}_{\rm e}$ as follows:
\begin{eqnarray} 
\lambda^{M}_{\perp,1} & = & \tfrac{1}{3} {\mathcal J}_{\rm Ising}+{\mathcal J}_{\rm iso} + {\mathcal J}_{\rm pd} \\
 \lambda^{M}_{\perp,2} & = & {\mathcal J}_{\rm iso} + \tfrac{1}{2}{\mathcal J}_{\rm pd} \\
 \lambda^{M}_{\parallel} & = & -\tfrac{2}{3} {\mathcal J}_{\rm Ising}+{\mathcal J}_{\rm iso} - 2 {\mathcal J}_{\rm pd}
\end{eqnarray}
This set of couplings does not include the Dzyaloshinskii-Moriya coupling, $\mathcal{J}_{\rm DM}$. 
Furthermore, in the fits to $\chiL$, Ref.~\cite{Malkin} imposes the constraint 
$\lambda^{M}_{\perp,1}=\lambda^{M}_{\perp,2}$. While this choice most likely circumvents 
the problem of having an underconstrained set of couplings to fit the $\chi_a$ data, the reduction of the 
parameter space to a particular two dimensional surface within a four dimensional space 
of couplings appears to us to be somewhat arbitrary. Indeed, the couplings we have obtained
 from neutron scattering data and which we used in our fit to $\chiL$ do not lie within 
the parameter space considered in Ref.~\cite{Malkin}. For example, $\mathcal{J}_{\rm DM}=-0.25$ K 
for our model with crystal field parameters taken from Ref.~\cite{Cao} (see Eq.~\ref{eqn:2}), 
which is of the same order of magnitude as the other three couplings. Reference \cite{Malkin} makes 
use of their model to study the exchange couplings of a number of rare earth pyrochlore titanates 
finding a significant degree of anisotropy in the exchange in a number of cases. 
For the case of $\YbTO$, which is relevant to this article, they mention couplings that are much 
larger than those of the other materials considered but do not report these couplings.



\Bibliography{30}

\bibitem{GGG} Gardner J S, Gingras M J P, and Greedan J E 2010 Magnetic pyrochlore oxides
{\it Rev. Mod. Phys.} {\bf 82}, 53.

\bibitem{Bramwell-Science} Bramwell S T and Gingras M J P 2001 Spin Ice State in Frustrated Magnetic Pyrochlore Materials {\it Science} {\bf 294}, 1495. 

\bibitem{Gardner-TTO-PRL}Gardner J S {\it et al.}
 2002 Co-operative Paramagnetism in the Geometrically Frustrated Pyrochlore, Tb$_2$Ti$_2$O$_7$
{\it Phys. Rev. Lett.} {\bf 82}, 1012.

\bibitem{Gingras-YMoO-PRL} Gingras M J P, Stager C V, Raju N P, Gaulin B D, and Greedan J E 1997 Static Critical Behavior of the Spin-Freezing Transition in the Geometrically Frustrated Pyrochlore Antiferromagnet Y$_2$Mo$_2$O$_7$  {\it Phys. Rev. Lett.} {\bf  78}, 947.

\bibitem{Gd2Sn2O7}Chapuis Y, Dalmas de R\'{e}otier P, Marin C, Yaouanc A, Forget A, Amato A, and Baines C 2009 Probing the ground state of Gd$_2$Sn$_2$O$_7$ through $\mu$SR measurements   {\it Physica B} {\bf 404} 686.

\bibitem{ETO} Lago J {\it et al.} 2005  Magnetic ordering and dynamics in the
  XY pyrochlore antiferromagnet: a muon-spin relaxation study of
  Er$_2$Ti$_2$O$_7$ and Er$_2$Sn$_2$O$_7$ {\it J. Phys.: Condens. Matter} {\bf
  17} 979.

\bibitem{111rod}Bonville P {\it et al.} 2004 Transitions and spin dynamics at very low temperature in the pyrochlores Yb$_2$Ti$_2$O$_7$ and Gd$_2$Sn$_2$O$_7$ {\it Hyperfine Interactions} {\bf 156}, 103. 

\bibitem{Cao}Cao H, Gukasov A, Mirebeau I, Bonville P, Decorse C, and Dhalenne G 2009 Ising versus XY Anisotropy in Frustrated R$_2$Ti$_2$O$_7$ Compounds as “Seen” by Polarized Neutrons {\it Phys. Rev. Lett.} {\bf 103}, 056402. 

\bibitem{Malkin}Malkin B Z, Lummen T T A, van Loosdrecht P H M, Dhalenne G, and Zakirov A R 2010 Static magnetic susceptibility, crystal field and exchange interactions in rare earth titanate pyrochlores {\it J. Phys.: Condens. Matter} {\bf 22}, 276003.


\bibitem{Onoda-1} Onoda S, and Tanaka Y 2011   Quantum fluctuations in the effective pseudospin-1/2 model for magnetic pyrochlore oxides arXiv:1011.4981 

\bibitem{Onoda-2} Onoda S  2010 Effective quantum pseudospin-1/2 model for Yb pyrochlore oxides arXiv:1101.1230

\bibitem{Hodges-YbTO-PRL} Hodges J A {\it et al.} 2002 First-Order transition in the spin dynamics of geometrically frustrated 
Yb$_{2}$Ti$_{2}$O$_{7}$ {\it Phys. Rev. Lett.} {\bf 88}, 077204. 


\bibitem{Yasui} Y. Yasui {\it et al.}, J. Phys. Soc. Jpn. {\bf 72}, 3014 (2003). 

\bibitem{Gardner-YbTO} Gardner J S, Ehlers G, Rosov N, Erwin R W, and Petrovic
  C 2004 Spin-spin correlations in Yb$_2$Ti$_2$O$_7$: A polarized neutron
  scattering study {\it Phys. Rev. B} {\bf 70}, 180404. 

\bibitem{Ross} Ross K A, Ruff J P, Adams C P, Gardner J S, Dabkowska H A, Qiu Y, Copley J R, Gaulin B D 2009 Two-dimensional kagome correlations and field induced order in the ferromagnetic XY pyrochlore Yb$_{2}$Ti$_{2}$O$_{7}$ {\it Phys. Rev. Lett.} {\bf 103} 227202.

\bibitem{YbTO} Thompson J D, McClarty P A, R\o{}nnow H M, Regnault L-P, Sorge A, and Gingras M J P 2010 Rods of Neutron Scattering Intensity in Yb$_2$Ti$_2$O$_7$: Compelling Evidence for Signifcant Anisotropic Exchange in a Magnetic Pyrochlore Oxide arXiv:1010.5476.

\bibitem{Hodges-YbTO-JPC} Hodges J A, Bonville P, Forget A, Rams M, Kr\'{o}las K, and Dhalenne G 2001 The crystal field and exchange interactions in Yb$_2$Ti$_2$O$_7$ {\it J. Phys. Condens. Matter} {\bf 13}, 9301. 

\bibitem{1742-6596-145-1-012032}McClarty P A, Curnoe, S H,  and Gingras M J P 2009 Energetic selection of ordered states in a model of the Er$_2$Ti$_2$O$_7$ frustrated pyrochlore XY antiferromagnet  {\it J. Phys.: Conference Series} {\bf 145}, 012032.

\bibitem{Canals2}Elhajal M, Canals B, Sunyer R, and Lacroix C 2005 Ordering in the pyrochlore antiferromagnet due to Dzyaloshinsky-Moriya interactions {\it Phys. Rev. B.} {\bf 71}, 094420.

\bibitem{Note} The measurements that have been made of the Curie-Weiss
  temperature \cite{Hodges-YbTO-JPC,Bramwell-JPC} all lie in the range $0.65\pm 0.15$ K.
  
 \bibitem{Onose} Onose Y, Ideue T, Katsura H, Shiomi Y, Nagaosa N, and Tokura Y 2010 Observation of the Magnon Hall Effect {\it Science} {\bf 329}, 297

\bibitem{Chern} Chern Gia-Wei, Fennie C J, and Tchernyshyov O 2006 Broken parity and a chiral ground state in the frustrated magnet CdCr$_{2}$O$_{4}$ {\it Phys. Rev. B} {\bf 74}, 060405



\bibitem{Gukasov} Gukasov A and Brown P J 2002 Determination of atomic site susceptibility tensors from polarized neutron diffraction data {\it J. Phys.: Condens. Matter} {\bf 14} 8831

\bibitem{Gingras-MFT}Enjalran M and Gingras M J P 2004 Theory of paramagnetic scattering in highly frustrated magnets with long-range dipole-dipole interactions: The case of the Tb$_2$Ti$_2$O$_7$ pyrochlore antiferromagnet {\it Phys. Rev. B} {\bf 70}, 174426.

\bibitem{Santini} Santini P {\it et al.} 2009 Multipolar interactions in f-electron systems: The paradigm of actinide dioxides {\it Rev. Mod. Phys.} {\bf 81} 807

\bibitem{Bramwell-JPC} Bramwell S T {\it et al.} 2000 {\it J. Phys.: Condens. Matter} {\bf 12} 483.










































\end{thebibliography}
\end{document}